\documentclass[a4paper,conference]{IEEEtran}
\usepackage[T1]{fontenc}
\usepackage{hyperref}
\usepackage{url}
\usepackage{booktabs} 
\usepackage{amsmath}    
\usepackage{amssymb}  
\usepackage{multirow}
\usepackage{placeins}
\usepackage{graphbox}
\usepackage{tikz}
\usepackage{float}

\usetikzlibrary{shapes.arrows}

\newcommand{\norm}[1]{\left\lVert#1\right\rVert}
\DeclareMathOperator*{\argmin}{arg\,min}

\definecolor{lightblue}{rgb}{0.85,0.88,0.91}
\definecolor{lightlightgray}{rgb}{0.85,0.85,0.85}

\makeatletter
\def\endthebibliography{%
	\def\@noitemerr{\@latex@warning{Empty `thebibliography' environment}}%
	\endlist
}
\makeatother

\ifCLASSINFOpdf
\else
\fi
\ifCLASSOPTIONcompsoc 
\usepackage[caption=false,font=normalsize,labelfont=sf,textfont=sf]{subfig} 
\else
\usepackage[caption=false,font=footnotesize]{subfig}
\fi
\begin{document}
%
\title{Phase Retrieval Using Conditional Generative Adversarial Networks}

\author{\IEEEauthorblockN{Tobias Uelwer, Alexander Oberstra\ss~and Stefan Harmeling}
\IEEEauthorblockA{Department of Computer Science\\
Heinrich Heine University D\"usseldorf\\
D\"usseldorf, Germany\\
Email: \{tobias.uelwer, alexander.oberstrass, stefan.harmeling\}@hhu.de}}


%


\maketitle

\begin{abstract}
\noindent In this paper, we propose the application of conditional generative adversarial networks to solve various phase retrieval problems.
We show that including knowledge of the measurement process at training time leads to an optimization at test time that is more robust to initialization than existing approaches involving generative models.
In addition, conditioning the generator network on the measurements enables us to achieve much more detailed results.
We empirically demonstrate that these advantages provide meaningful solutions to the Fourier and the compressive phase retrieval problem and that our method outperforms well-established projection-based methods as well as existing methods that are based on neural networks.
Like other deep learning methods, our approach is robust to noise and can therefore be useful for real-world applications.
\end{abstract}


%
\IEEEpeerreviewmaketitle

\section{Introduction}
Phase retrieval is an important problem which has applications e.g. in X-ray crystallography \cite{millane1990phase}, astronomical imaging \cite{fienup1987phase}, microscopy \cite{zheng2013wide} and many more.
For the sake of readability we only define the phase retrieval problems  for one-dimensional signals. 
The extension to higher dimensions is straight forward.
\subsection{The Compressive Phase Retrieval Problem}
The compressive phase retrieval problem \cite{compressive} can be defined as recovering a signal $x\in \mathbb{R}^n$ given $m$ measurements $y\in \mathbb{R}^m$, where the dependence of the measurements on the signal is described by 
\begin{equation}
\label{eq1}
y:=|Ax| \text{ for } A\in \mathbb{C}^{m\times n}.
\end{equation}  
The matrix $A$ is usually called the measurement matrix. 
For $m< n$ the problem is also called compressive phase retrieval.

\subsection{The Fourier Phase Retrieval Problem}
Having $A$ as the symmetric Fourier transformation matrix $\mathcal{F}$, containing the primitive roots of unity, results in the Fourier phase retrieval problem, which thus can be seen as a special case of the compressive phase retrieval problem.
Since the Fourier phase retrieval problem is very relevant in practice, we want to focus on this problem.
Both phase retrieval problems are highly ill-posed.
As a result, prior knowledge about the signal $x$ is necessary for the reconstruction process.

\subsection{Prior Work}

\begin{figure*}
	\centering
	\begin{tikzpicture}
	\node[inner sep=0pt] (orig) at (0,0) {\includegraphics[width=.06\textwidth]{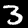}};
	\node[] (x) at (0,-1.2) {$x\in \mathbb{R}^{28\times 28}$};
	
	\node[inner sep=0pt] (orig) at (5,0) {\includegraphics[width=.06\textwidth]{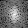}};
	\node[] (y) at (5,-1.2) {$y\in \mathbb{R}^{28\times 28}$};
	
	\node[inner sep=0pt] (orig) at (10,0) {\includegraphics[width=.06\textwidth]{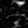}};
	\node[] (xhat) at (10,-1.2) {$\hat{x}\in \mathbb{R}^{28\times 28}$};
	
	\node[inner sep=0pt] (orig) at (0,-3.5) {\includegraphics[width=.12\textwidth]{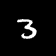}};
	\node[] (x2) at (0,-4.9) {$x\in \mathbb{R}^{56\times 56}$};
	
	\node[inner sep=0pt] (orig) at (5,-3.5) {\includegraphics[width=.12\textwidth]{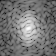}};
	\node[] (y2) at (5,-4.9) {$y\in \mathbb{R}^{56\times 56}$};
	
	\node[inner sep=0pt] (orig) at (10,-3.5) {\includegraphics[width=.12\textwidth]{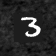}};
	\node[] (xhat2) at (10,-4.9) {$\hat{x}\in \mathbb{R}^{56\times 56}$};

	\node[draw, single arrow,
	minimum height=18mm, minimum width=4mm,
	single arrow head extend=2mm,
	anchor=center] at (2.5,0) {};
	\node[draw, single arrow,
	minimum height=18mm, minimum width=4mm,
	single arrow head extend=2mm,
	anchor=center] at (7.5,0) {};
	\node[draw, single arrow,
	minimum height=18mm, minimum width=4mm,
	single arrow head extend=2mm,
	anchor=center] at (2.5,-3.5) {};
	\node[draw, single arrow,
	minimum height=18mm, minimum width=4mm,
	single arrow head extend=2mm,
	anchor=center] at (7.5,-3.5) {};
	
	\node[] (m1) at (2.5,-0.7) {Measurement};
	\node[] (m2) at (7.5,-0.7) {HIO};
	\node[] (hio2) at (2.5,-4.2) {Measurement};
	\node[] (hio2) at (7.5,-4.2) {HIO};
	\end{tikzpicture}
	\caption{Reconstructing a signal with a known zero padding leading to oversampling of the measurement is a lot easier. However without padding the commonly used HIO algorithm fails to reconstruct the image.}
	\label{fig:hio-pad}
\end{figure*}
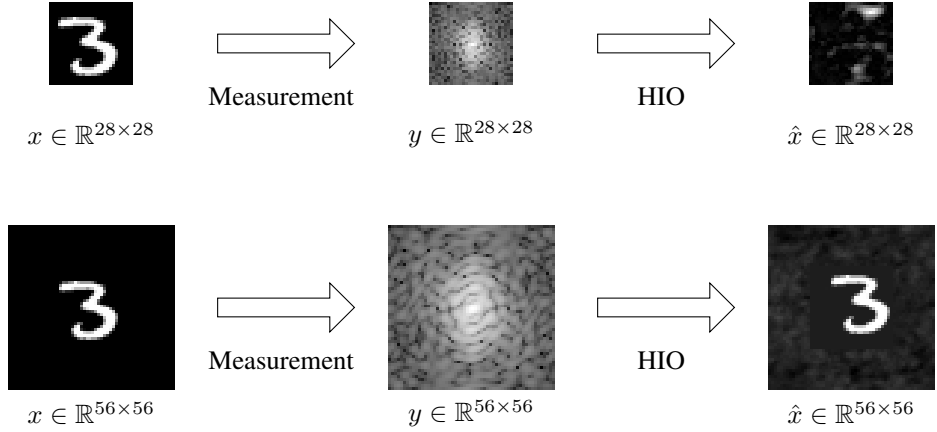

In practice, traditional projection-based methods like the Gerchberg-Saxton algorithm \cite{gerchberg}, the hybrid input-output (HIO) algorithm \cite{fienup} or the relaxed averaged alternating reflections (RAAR) algorithm \cite{raar} require the magnitudes of the Fourier measurements to be oversampled \cite{iterative}, as demonstrated in Figure \ref{fig:hio-pad}.
The influence of oversampling on the Fourier phase retrieval problem is discussed in \cite{miao}.
Without oversampling more prior knowledge about the signal is needed.
Traditional methods, however, merely take some handcrafted prior information like signal sparsity into the reconstruction process. 
Sparse phase retrieval has been discussed in \cite{jaganathan2013sparse, tillmann2016dolphin, cai2016optimal, elser2018benchmark}. 
Neural networks allow the inclusion of self-learned prior knowledge and thus can be used to remedy this problem.
Phase retrieval using generative models has been previously suggested by Hand et al. \cite{generative_prior}. 
We describe their approach briefly in Section \ref{sec:generative-prior}.
Furthermore, neural networks have been used to improve the results of existing approaches: Metzler et al. \cite{prdeep} apply the regularization-by-denoising framework to the phase retrieval problem, where a denoising neural networks is used to construct a regularization term. 
Also I{\c{s}}{\i}l et al. \cite{icsil2019deep} used neural networks to remove artifacts that are produced during the iterations of the HIO algorithm.
Both of these apporoaches were only introduced for the oversampled case.
End-to-end learning for phase retrieval has been discussed by Nishizaki et al. \cite{nishizaki2020analysis}.
Conditional generative adversarial networks (conditional GANs) have previously been applied to Fourier ptychography by Boominathan et al. \cite{cganptych}. Fourier ptychography aims to combine multiple phaseless measurement vectors $y_1, \dots, y_k$ into a single reconstruction. The problem is significantly different from the more generic problem we want to solve. The relationship between theses problems is discussed in \cite{acceleratedwf}.

\section{Methods}
\label{sec:methods}
\subsection{End-to-End Learning (E2E)}
\label{sec:end-to-end}
As a baseline we use a neural network generator $G$, trained to directly output an approximate reconstruction $\hat{x}$ of the input signal $x$ by the magnitudes $y$:
\begin{equation}
\label{e2e}
\hat{x} = G(y)
\end{equation}

In training, we optimize the distance between the original signal $x$ and the reconstruction $\hat{x}$, where we construct the input measurements $y$ by applying Equation \ref{eq1} on the test samples $x$. 
When optimizing the Euclidean distance the reconstructed images tend to be blurry.
A similar observation has been made in the context of image inpainting by Pathak el al. \cite{context}.
Therefore, we use the mean absolute error for training, which produces slightly less blurry reconstructions.
We refer to the method as E2E.

\subsection{Phase Retrieval Using a Generative Prior (DPR)}
\label{sec:generative-prior}
Instead of directly performing optimization on the input signal $x$ itself, which was for example done by Candes et al. \cite{candes2015phase}, Hand et al. \cite{generative_prior} suggest viewing the problem through the lens of a generative model $G$, i.e., by optimizing its latent variable 
\begin{equation}
\label{eq:dpr-min}
z^*=\argmin_z \norm{y-|AG(z)|}^2_2.
\end{equation}
This allows incorporating the prior knowledge of the generative model $G$ that has been previously trained on data similar to the signal of interest.
This optimization problem is solved using a modified gradient descent algorithm that accounts for solutions corresponding to latent variables having a flipped sign. 
Furthermore, the optimization result strongly depends on the initialization of the latent variable $z$ and often gets stuck in local minima, as we observed for the Fourier phase retrieval problem.
In practice, we found that the dimension of the latent variable $\dim(z)$ must be chosen sufficiently small to reduce the distance $\norm{y-|AG(z)|}^2_2$ far enough.
Finding an optimal point $z^*$ with Equation \ref{eq:dpr-min} results in the estimated signal $\hat{x}$ by 
\begin{equation}
\label{dpr}
\hat{x}=G(z^*).
\end{equation}
Throughout the paper we denote this method as DPR.

\subsection{Phase Retrieval using Conditional Generative Adversarial Networks (PRCGAN)}\label{sec:prcgan}
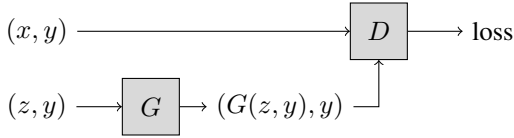
\begin{figure}
	\centering
	\begin{tikzpicture}
		\node[rectangle, fill=lightlightgray, minimum height=0.75cm,minimum width=0.75cm,draw] (G) at (-1.5, 0) {\(G\)};
		\node[rectangle, fill=lightlightgray, minimum height=0.75cm,minimum width=0.75cm,draw] (D) at (1.5, 1) {\(D\)};
		\node[] (x) at (-3,1) {$(x,y)$};
		\node[] (z) at (-3,0) {$(z,y)$};
		\node[] (genout) at (0.2,0) {$(G(z,y),y)$};
		\node[] (loss) at (3,1) {loss};
		\path[draw, ->] (genout) -| (D);
		\draw[->] (G) edge (genout);
		\draw[->] (x) edge (D);
		\draw[->] (z) edge (G);
		\draw[->] (D) edge (loss);
	\end{tikzpicture}
	\caption{Overview of the conditional generative adversarial network approach.}
	\label{fig:cgan}
\end{figure}

In this work we suggest to use a conditional GAN approach which was proposed by Goodfellow et al. \cite{gan} and Mirza and Osindero \cite{conditional_gan}. 
Furthermore, we propose to optimize the latent variable after training to minimize the measurement error.
Our approach can be seen as a hybrid of the methods described in Section \ref{sec:end-to-end} and Section \ref{sec:generative-prior}.
In the following, we denote the distribution of the latent variable $z$ by $q$, whereas we denote the unknown data distribution by $p$.
Although the latent distribution can be chosen arbitrarily, we use the standard normal distribution which is a common choice for GAN training.
We chose the dimension of the latent variable $z$ to be equal to the dimension of the measurement $y$.
The training objective for the generator $G$ and the discriminator $D$ consists of an adversarial component 
\begin{equation}
\label{eq:adv}
\begin{split}
\mathcal{L}_{\text{adv}}(D,G)&=\mathbb{E}_{x\sim p} \big[\log D(x,y)\big] \\
&+\mathbb{E}_{x\sim p,z\sim q}  \big[\log \big(1-D(G(z, y),y)\big)\big], 
\end{split}
\end{equation} where
\begin{equation}
y:=|Ax|
\end{equation} 
as given by Equation \ref{eq1} and a reconstruction component 
\begin{equation}
\mathcal{L}_{\text{rec}}(G) = \mathbb{E}_{x\sim p,z\sim q} \big[ \norm{x-G(z,y)}_1\big]. 
\end{equation}
The adversarial component encourages the generator to output realistic and sharp images, whereas the reconstruction guides the generator in the right direction. 
The optimization problem that is solved during training is given by
\begin{equation}
\min_G  \max_D \mathcal{L}_{\text{adv}}(D,G)+ \lambda \mathcal{L}_{\text{rec}}(G),
\end{equation}
where the hyperparameter $\lambda$ can be used to control the influence of both losses. 
The discriminator network $D$ gets the original image (or the reconstruction) as well as the measurements as input.
Figure \ref{fig:cgan} gives an overview of our approach.
Due to stability issues we also used the modified GAN loss that was suggested by Goodfellow et al. \cite{gan}.
We also trained the conditional GAN using the least squares objective proposed by Mao et al. \cite{mao2017least}, but we did not find the results to be better than with the logarithmic loss function as stated in Equation \ref{eq:adv}.
Samples generated by 
\begin{equation}
\label{prcgan}
\hat{x} = G(z, y),\text{ with } z \sim \mathcal{N}(0,1)
\end{equation}
already lead to decent reconstructions $\hat{x}$ of the original signal $x$.
However, to further improve the results at test time, we adapt the method from Hand et al. \cite{generative_prior} to seek for the optimal latent variable $z^*$ for each data point that minimizes the squared Euclidean error of the measurements:
\begin{equation}
\label{cgan-opt}
z^* = \argmin_z \norm{y-|AG(z, y)|}^2_2.
\end{equation}
After the optimization of the latent variable $z$, the estimate of the signal is given by
\begin{equation}
\label{prcganstar}
\hat{x}=G(z^*, y).
\end{equation}
During training, the network learns small values for the filter of the latent variable, i.e., it learns to ignore the noise. 
This requires us to employ large learning rates for the latent optimization during test time.
Mathieu et al. \cite{beyond_mse} and Isola et al. \cite{image_translation} made a similar observation and removed the latent variable. 
Isola et al. \cite{image_translation} introduced additional stochasticity into the model by applying dropout during training time and testing time.
However, none of these approaches yielded better results in our experiments.
We also experimented with dropping the latent variable and training an end-to-end network with adversarial loss, but the results were worse.
In the following we refer to our approach without latent optimization as PRCGAN.
Furthermore, we refer to the PRCGAN with latent space optimization as PRCGAN*.
We show that our PRCGAN* consistently produces better results than PRCGAN.

\section{Experiments: Fourier Phase Retrieval}
In this section, we empirically rate the quality of the predictions generated by the proposed method from Section \ref{sec:prcgan}.

\subsection{Datasets} 
Commonly used datasets in deep learning are MNIST \cite{mnist}, Fashion-MNIST \cite{fashion} and CelebA \cite{celeba}. 
Images from MNIST and Fashion-MNIST both have the same shape of $28 \times 28$ pixels with only one greyscale channel. 
They both provide $60\,000$ training and $10\,000$ test samples. 
For computational reasons, we reduced the test size to $1024$ samples. 
Fashion-MNIST has more variance in the image representations than MNIST which makes it harder to memorize significant patterns for our generative models.
Therefore, we consider it as a harder dataset for phase retrieval than MNIST.
As a third dataset, we took the popular CelebA dataset \cite{celeba}, which consists of $202\,599$ images of human faces.
We took the first $162\,769$ images as training data, the next $19\,867$ as validation data and from the remaining images we took the first $1024$ for evaluation.
To reduce the size of the high resolution images to more computational friendly sizes, we took a center crop of $108 \times 108$ pixels from each image and then resized it to a total scale of $64 \times 64$ pixels.
This transformation is the same that was used by Hand et al. \cite{generative_prior}.
One difference of the CelebA dataset to the previously introduced datasets is that it consists of $3$ color channels.
During the measurement process, we treated each channel independently and, in the case of Fourier measurements, we performed a 2D Fourier transform for each channel separately.

\subsection{Evaluation} 
To evaluate the performance of the models, we compared the distance of the reconstruction $\hat{x}$ to the original signal $x$.
For this we use the mean squared error (MSE), mean absolute error (MAE) and structural similarity (SSIM) \cite{ssim}. 
These metrics do not take visual aspects like sharpness into account \cite{wang2009mean}.
Especially the mean squared error tends to prefer blurry predictions. Since the Fourier measurements are invariant under signal translation and rotation by $180$ degrees, we sometimes observed these transformations in our reconstruction results as well. 
Although we consider these reconstructions as equally correct solutions, pixel-wise metrics like MSE do not take this property of the Fourier transformation into account.
We therefore used a cross-correlation based image registration technique \cite{phasebased} to estimate the most probable translation by the predicted signal $\hat{x}$ relative to the original signal $x$.
We also calculated the optimal translation for the image rotated by 180 degree and we reported the result with minimal error of these two.
For MNIST and Fashion-MNIST we registered the predictions before calculating the evaluation metrics.
For the CelebA dataset we omitted the registration since we did not observe any effect.

\subsection{MNIST and Fashion-MNIST}
\label{MNIST-Fashion}
\begin{table*}
	\centering
	\caption{Evaluation results for MNIST, Fashion-MNIST and CelebA for the reconstructions from the Fourier magnitudes. We register the reconstructions for MNIST and Fashion-MNIST. MSE, MAE: lower is better. SSIM: higher is better.}
	\label{results}	
	\begin{tabular}{llcccccc}
		\toprule
		Dataset & Metric & HIO & RAAR & E2E & DPR & PRCGAN & PRCGAN*  \\
		\midrule
		\multirow{3}{*}{MNIST}
		& MSE & $0.0441$ & $0.0489$ & $0.0183$ & $0.0093$ & $0.0168$ & $\mathbf{0.0010}$ \\
		& MAE & $0.1016$ & $0.1150$ & $0.0411$ & $0.0221$ & $0.0399$ & $\mathbf{0.0043}$ \\
		& SSIM & $0.5708$ & $0.5232$ & $0.8345$ & $0.9188$ & $0.8449$ & $\mathbf{0.9898}$ \\
		\midrule
		\multirow{3}{*}{Fashion-MNIST}
		& MSE & $0.0646$ & $0.0669$ & $0.0128$ & $0.0280$ & $0.0151$ & $\mathbf{0.0087}$ \\
		& MAE & $0.1604$ & $0.1673$ & $0.0526$ & $0.0856$ & $0.0572$ & $\mathbf{0.0412}$ \\
		& SSIM & $0.4404$ & $0.4314$ & $0.7940$ & $0.6602$ & $0.7749$ & $\mathbf{0.8580}$ \\
		\midrule
		\multirow{3}{*}{CelebA}
		& MSE & $0.0737$ & $0.0729$ & $0.0106$ & $0.0388$ & $0.0138$ & $\mathbf{0.0093}$ \\
		& MAE & $0.2088$ & $0.2073$ & $0.0699$ & $0.1323$ & $0.0804$ & $\mathbf{0.0642}$ \\
		& SSIM & $0.1671$ & $0.2274$ & $0.7444$ & $0.5299$ & $0.6799$ & $\mathbf{0.7631}$ \\
		\bottomrule
	\end{tabular}
\end{table*}

\subsubsection{HIO} 
We ran $1000$ iterations of the HIO algorithm, where we set the hyperparameter $\beta=0.8$ with $3$ random restarts.
We did not zero-pad the original signal and reported the test error of the reconstruction $\hat{x}$ with the lowest measurement error $\norm{|A\hat{x}|-y}_2$.

\subsubsection{RAAR} 
We ran 1000 iterations of the RAAR algorithm with $\beta=0.87$, as it was reported to be the best choice for $\beta$ by Luke \cite{raar}.
To overcome very bad starting points, we took the best out of three random initializations as for the HIO.

\subsubsection{End-to-End} 
For the E2E approach we used a generator $G(y)$, where $y$ denotes the measurement information, with 5 fully connected layers of sizes $784 - 2048 - 2048 - 2048 -  2048 - 784$ for both datasets.
We preferred using fully-connected layers over convolutional layers to match the property of each measurement depending on every image pixel and the other way around.
This assumption was consistent with our observation that fully-connected networks performed slightly better over pure convolutional networks in our experiments.
As input at train and test time, we took the measurement information $y$ for each data sample by applying Equation \ref{eq1} and flattened it to a vector of size $784$ to feed the first fully-connected layer.
This first layer then mapped the input vector to a hidden size of $2048$. 
The last layer mapped the $2048$ hidden values back again to $784$ pixel values that were then rearranged to a $28 \times 28$ image output again. 
We placed batch-normalization \cite{batchnorm} and ReLU activation functions in between all fully-connected layers.
The output of the last layer was then passed to a Sigmoid function to ensure the output was in the range $[0, 1]$. 

\subsubsection{DPR}
As generative model $G$ for the generative prior approach we used a Variational Autoencoder (VAE) \cite{vae} similar to the one proposed by Hand et al. \cite{generative_prior}, but we had more success with a higher dimension of ${128}$ for of the latent space $z$ in the Fourier measurement case. 
Therefore, our decoder network got the size of $128 - 500 - 500 - 784$ (with the encoder vice versa).
While this choice of model size resulted in 10 times fewer learnable parameters than for the comparable E2E and PRCGAN approaches, we did not observe any improvement in performance when using a larger model or using a GAN instead of VAE.
After training the VAE, we performed the optimization steps as described in Section \ref{sec:generative-prior} for a random initialization $z \sim \mathcal{N}(0, 1)$ to find an optimal $z^*$ and recorded $G(z^*)$.
We ran $10\,000$ optimization steps with a learning rate of $0.1$ and the Adam optimizer \cite{kingma2014adam}, which we found performed best.
Since the optimization often got stuck in local minima depending on the initialization of the latent space, we recorded only the best out of $3$ random restarts.

\subsubsection{PRCGAN}
To keep the approaches comparable, we took the same model structure as for the E2E case but with a twice as large input size for the additional latent noise $z$, resulting in a generator $G$ with sizes $1568 - 2048 - 2048 - 2048 -  2048 - 784$.
During training we found the choice of hyperparameter $\lambda = 1000$ performed best.
Lower values resulted in artifacts like scattered dots throughout the image.
Larger values led to blurred results, which is consistent with the outcomes of the E2E approach.
The output of PRCGAN, based only on the conditional information with a random latent noise initialization of $z$, already produced reasonable results.
We then performed the same amount of optimization steps for our PRCGAN* approach as for the DPR approach to solve Equation \ref{cgan-opt} to find an optimal latent variable $z^*$.
However, we were required to use a much larger learning rate for this model.
This is caused by the small weights for latent variable learned during training. 
The large learning rate overcame this problem and produced even better results with the optimized latent variable $z^*$ than the results with random $z$.

\subsubsection{Results} 
\begin{figure}
	\centering
	\begin{tabular}{rl} 
		Original & \includegraphics[height=0.72cm, align=c]{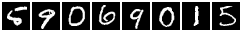}\\
		\vspace{0.01cm}\\
		HIO \cite{fienup}& \includegraphics[height=0.72cm, align=c]{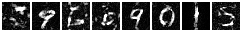}\\
		\vspace{0.01cm}\\
		RAAR \cite{raar}& \includegraphics[height=0.72cm, align=c]{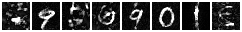}\\
		\vspace{0.01cm}\\
		E2E  & \includegraphics[height=0.72cm, align=c]{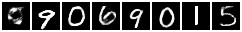}\\
		\vspace{0.01cm}\\
		DPR \cite{generative_prior} & \includegraphics[height=0.72cm, align=c]{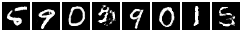}\\
		\vspace{0.01cm}\\
		PRCGAN (ours) & \includegraphics[height=0.72cm, align=c]{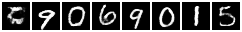}\\
		\vspace{0.01cm}\\
		PRCGAN* (ours) & \includegraphics[height=0.72cm, align=c]{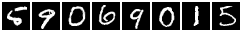}\\
		\vspace{0.01cm}\\
	\end{tabular}
	\caption{Registered reconstructions from the Fourier magnitudes of samples from the MNIST test dataset (not cherry-picked) for each model.}
	\label{plots-MNIST}
\end{figure}

\begin{figure}
	\centering
	\begin{tabular}{rl} 
		Original & \includegraphics[height=0.72cm, align=c]{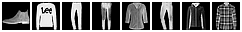}\\
		\vspace{0.01cm}\\
		HIO \cite{fienup} & \includegraphics[height=0.72cm, align=c]{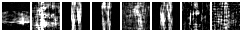}\\
		\vspace{0.01cm}\\
		RAAR \cite{raar}& \includegraphics[height=0.72cm, align=c]{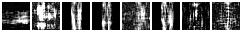}\\
		\vspace{0.01cm}\\
		E2E & \includegraphics[height=0.72cm, align=c]{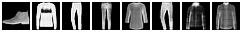}\\
		\vspace{0.01cm}\\
		DPR \cite{generative_prior}  & \includegraphics[height=0.72cm, align=c]{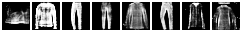}\\
		\vspace{0.01cm}\\
		PRCGAN (ours) & \includegraphics[height=0.72cm, align=c]{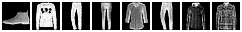}\\
		\vspace{0.01cm}\\
		PRCGAN* (ours) & \includegraphics[height=0.72cm, align=c]{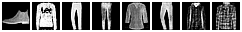}\\
		\vspace{0.01cm}\\
	\end{tabular}
	\caption{Registered reconstructions from the Fourier magnitudes of samples from the Fashion-MNIST test dataset (not cherry-picked) for each model.}
	\label{plots-Fashion}
\end{figure}

Table \ref{results} summarizes the evaluation results for various metrics on the full $1024$ sample test sets.
Figures \ref{plots-MNIST} and \ref{plots-Fashion} show a comparison of the reconstruction of the first eight test samples from the MNIST and Fashion-MNIST datasets.
The first lines in Figure \ref{plots-MNIST} and \ref{plots-Fashion} show the original ground truth.
Traditional methods like HIO and RAAR were not able to converge to the correct solution, resulting in fragmented, blurry outputs on both datasets.
While the E2E approach overcame the problem of constructing fragmented parts, it still produced blurry outputs.
Due to the adversarial loss component, the PRCGAN approach painted finer texture components like the text on the sweatshirt in the second sample and the check pattern on the shirt in the eighth sample shown in Figure \ref{plots-Fashion}.
While on MNIST the numerical results are consistent with this observation, our evaluation metrics rate the E2E better in the case of Fashion-MNIST than the PRCGAN as shown in Table \ref{results} because they do not take these visual aspects into account and prefer a risk minimized output without the risk of misplaced sharp edges.
On MNIST, the approaches with latent optimization, DPR and PRCGAN*, produce the best visual appearance of the digits among all approaches.
However, DPR gets stuck in local minima sometimes, as one can see from the fourth and eighth test sample in Figure \ref{plots-MNIST}.
With other random initializations these samples got reconstructed perfectly, but we only allowed $3$ random restarts to keep the computational effort limited.
The numerical results in Table \ref{results} also show that DPR and PRCGAN* perform best on MNIST, while the PRCGAN* performed even better because it is not affected by local minima in the optimization landscape.
On the Fashion dataset, the rating of DPR dropped dramatically.
The outputs become even more random than on MNIST.
While PRCGAN* still performed best, its advantage over the other methods became smaller.
However, it is remarkable that this is the only approach that produced a readable variant of the text on the sweatshirt in the second sample in Figure \ref{plots-Fashion}.

\begin{figure}
	\centering
	\subfloat[MNIST]{\includegraphics[width=0.45\textwidth]{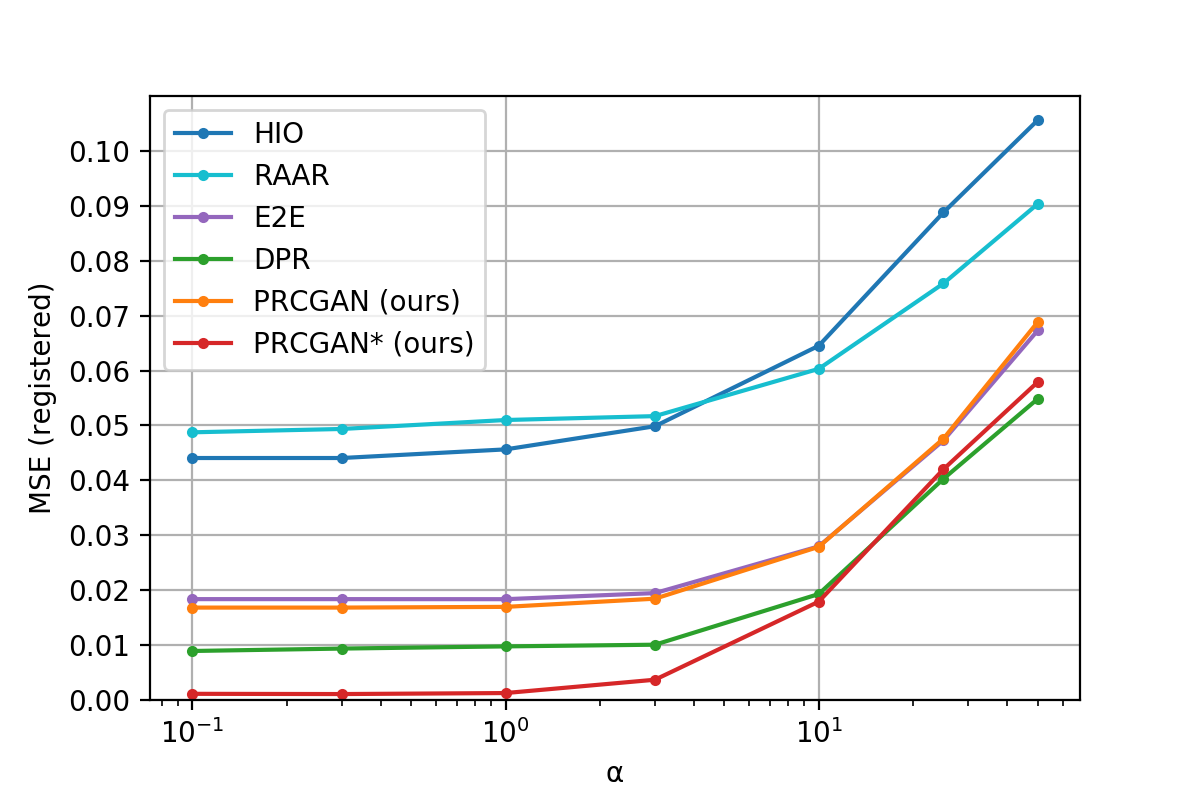}}\\
	\subfloat[Fashion-MNIST]{\includegraphics[width=0.45\textwidth]{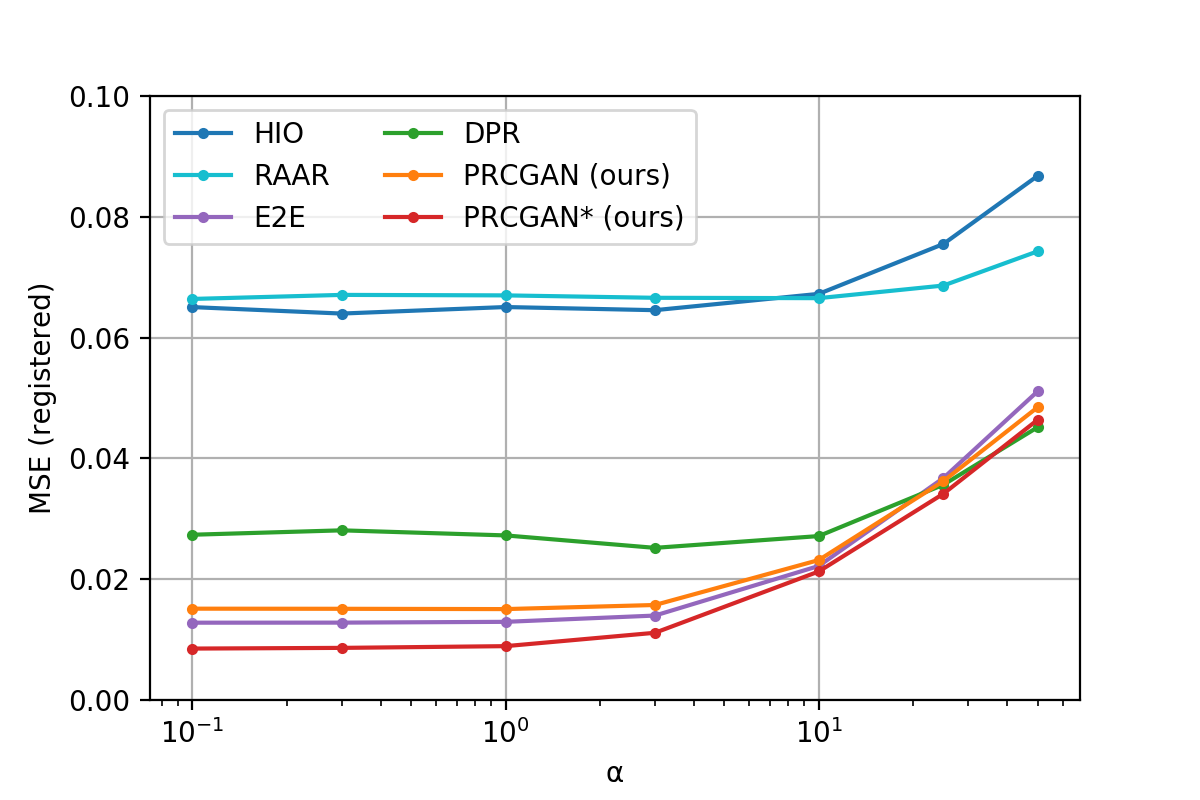}}
	\caption{MSE of registered reconstructions from noisy Fourier magnitudes  from the MNIST and Fashion-MNIST datasets. $\alpha$ controls the noise level. Higher $\alpha$ corresponds to stronger noise.}
	\label{fig:noise-plots}
\end{figure}

\paragraph{Robustness to Noise}

To train our PRCGAN, we create synthetic noiseless measurements.
However, real experiments contain several sources of noise that can disrupt the reconstruction process.
In this section, we show that our model, as other deep learning approaches, is still very robust to noise. 
Experiments often measure intensities as the squared Fourier magnitudes through discrete photon counting, as performed in X-ray crystallography \cite{elser2018benchmark}, resulting in additive noise
\begin{equation}
\hat{y}^2 = y^2 + w
\end{equation} for the measured intensities $\hat{y}^2$ consisting of the true intensities $y^2$ and noise $w$.
Shot noise is one of the dominant sources of noise for photon counts \cite{yeh2015experimental} and is used by Metzler et al. \cite{prdeep} to show robustness to noise.
They suggest attaining the noisy magnitudes $\hat{y}$ by sampling
\begin{equation}
\begin{split}
s &\sim \text{Poisson}\left(\frac{y^2}{\alpha^2}\right)\\
\hat{y} &= \alpha\sqrt{s}
\end{split}
\end{equation}
where $\alpha$ controls the variance of the random variable $\hat{y}^2$ and therefore the signal to noise ratio (SNR).
However, the variance additionally depends on the scale of the true magnitudes $y$ that can differ between experimental setups.
Therefore, we also measured the SNR of the noisy model input $\hat{y}$ directly for each sample by
\begin{equation}
\operatorname{SNR} = \frac{\mu_\mathrm{magn}}{\sigma_\mathrm{noise}}
\end{equation}
where $\mu_\mathrm{magn}$ denotes the mean of the true magnitudes $y$ and $\sigma_\mathrm{noise}$ the standard deviation of the error $\hat{y} - y$.
Figure \ref{fig:noise-plots} shows the reconstruction results for the proposed methods as the mean MSE for different values of $\alpha$ on the MNIST and Fashion-MNIST datasets for each $1024$ samples.
Both plots show that the proposed method is robust to noise up to $\alpha = 3$ which lead to a $\text{SNR}$  of approximately $3$.
All deep learning approaches show similar robustness to noise.
The slight leading and unexpected observed improvement of the DPR in the high variance regions we attribute to the $3$ allowed random restarts that increase their impact the more random the outputs get.

\subsection{CelebA}

\begin{figure}
	\centering
	\begin{tabular}{rl} 
		Original & \includegraphics[height=0.65cm, align=c]{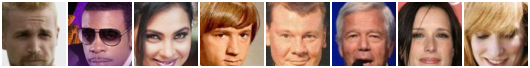}\\
		\vspace{0.01cm}\\
		HIO \cite{fienup} & \includegraphics[height=0.65cm, align=c]{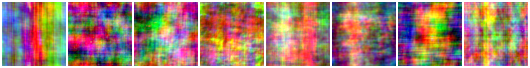}\\
		\vspace{0.01cm}\\
		RAAR \cite{raar} & \includegraphics[height=0.65cm, align=c]{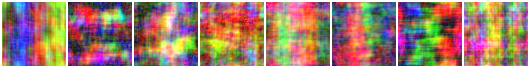}\\
		\vspace{0.01cm}\\
		E2E & \includegraphics[height=0.65cm, align=c]{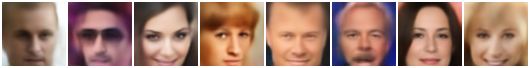}\\
		\vspace{0.01cm}\\
		DPR \cite{generative_prior}  & \includegraphics[height=0.65cm, align=c]{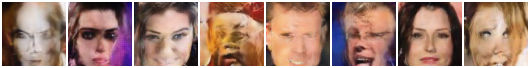}\\
		\vspace{0.01cm}\\
		PRCGAN (ours) & \includegraphics[height=0.65cm, align=c]{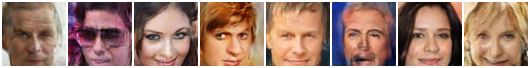}\\
		\vspace{0.01cm}\\
		PRCGAN* (ours) & \includegraphics[height=0.65cm, align=c]{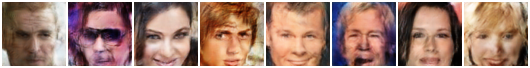}\\
		\vspace{0.01cm}\\
	\end{tabular}
	\caption{Reconstructions of samples from the CelebA test dataset (not cherry-picked) for each model.}
	\label{plots-celeba}
\end{figure}

\subsubsection{E2E}
The flattened version of images from CelebA, as we did for the fully-connected models for MNIST and Fashion-MNIST, would result in a $12\,288$ dimensional input vector for the CelebA dataset.
Since this would lead to very large matrices within the fully-connected layers, they were not suitable enough anymore.
To still account for the global dependence of each measurement on the signal, we first compressed the input for the E2E model by $5$ convolutional layers.
After the last convolutional layer the input had a flattened size of $2048$.
We then were able to append $2$ fully-connected layers with moderate computational effort, preserving the dimension of $2048$.
The output of the fully-connected layers was then upscaled to the full image size by $5$ transposed convolutional layers again.
The output of the last layers was then passed to a Sigmoid function to ensure the output of the network was in the range $[0, 1]$.
We used batch-normalization and ReLU activation functions between all layers.

\subsubsection{DPR}
As a generative model we used a DCGAN \cite{radford2015unsupervised} which we trained for $100$ epochs.
Similar to the DPR optimization for MNIST and Fashion-MNIST, we ran $10\,000$ optimization steps with a learning rate of $0.1$ and the Adam optimizer to find an optimal latent variable $z^*$.
Again, we gave the DPR approach $3$ restarts for different random initializations and only recorded the best result.

\subsubsection{PRCGAN}
To get comparable results, we used the same model structure as for the E2E approach with additional $3$ channels for an additional $3 \times 64 \times 64$ dimensional latent input $z$ to the measurement information $y$ of the same shape.
Again, setting $\lambda = 1000$ worked best, while lower values caused some random artifacts and higher values caused increasing blurriness.
We did the same latent optimization as described for MNIST and Fashion-MNIST for this approach on CelebA as well.

\subsubsection{Results}
Table \ref{results} contains the numerical results for various metrics by the different approaches for CelebA.
The plots in Figure \ref{plots-celeba} again show the reconstruction for 8 test samples, where the first row contains the original ground truth.
The traditional approaches like HIO and RAAR were completely overwhelmed by this task and in contrast to MNIST and Fashion-MNIST both methods did not produce any recognizable patterns anymore.
While the DPR approach only got stuck sometimes in the optimization process with samples from MNIST and Fashion-MNIST, it produced very distorted outputs almost all the time.
The PRCGAN* still performed best out of all approaches, although the lead over the E2E and PRCGAN approaches became smaller.
While E2E outperformed PRCGAN in all evaluation metrics, it produced very blurry outputs.
The human eye can easily distinguish between the original and generated images from E2E.
The PRCGAN generated much more natural looking outputs and preserved the edges contained in the image.

\section{Experiments: Compressive Phase Retrieval}

\begin{figure*}[]
	\centering
	\begin{tabular}{ccccccccc|c}
		E2E & 
		\includegraphics[width=0.8cm, align=c]{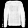} &
		\includegraphics[width=0.8cm, align=c]{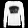} & 
		\includegraphics[width=0.8cm, align=c]{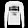} & 
		\includegraphics[width=0.8cm, align=c]{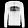} & 
		\includegraphics[width=0.8cm, align=c]{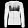} & 
		\includegraphics[width=0.8cm, align=c]{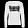} & 
		\includegraphics[width=0.8cm, align=c]{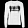} & 
		\includegraphics[width=0.8cm, align=c]{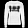} &
		\includegraphics[width=0.8cm, align=c]{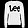} \vspace{2mm}\\
		DPR \cite{generative_prior} & 
		\includegraphics[width=0.8cm, align=c]{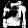} &
		\includegraphics[width=0.8cm, align=c]{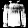} & 
		\includegraphics[width=0.8cm, align=c]{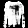} & 
		\includegraphics[width=0.8cm, align=c]{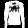} & 
		\includegraphics[width=0.8cm, align=c]{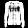} & 
		\includegraphics[width=0.8cm, align=c]{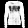} & 
		\includegraphics[width=0.8cm, align=c]{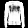} & 
		\includegraphics[width=0.8cm, align=c]{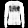} &
		\includegraphics[width=0.8cm, align=c]{figures/results_Ameas/Fashion/original.png} \vspace{2mm}\\
		PRCGAN* (ours) & 
		\includegraphics[width=0.8cm, align=c]{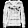} &
		\includegraphics[width=0.8cm, align=c]{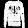} & 
		\includegraphics[width=0.8cm, align=c]{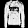} & 
		\includegraphics[width=0.8cm, align=c]{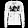} & 
		\includegraphics[width=0.8cm, align=c]{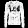} & 
		\includegraphics[width=0.8cm, align=c]{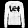} & 
		\includegraphics[width=0.8cm, align=c]{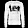} & 
		\includegraphics[width=0.8cm, align=c]{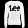} &
		\includegraphics[width=0.8cm, align=c]{figures/results_Ameas/Fashion/original.png} \vspace{2mm}\\
		m & 10 & 25 & 50 & 100 & 200 & 300 & 500 &784& True
	\end{tabular}
	\caption{Reconstructions of an Fashion-MNIST test image for a varying number of measurements $m$.}
	\label{fig:gaussian_measurements}
\end{figure*}

\begin{figure}[H]
	\centering
	\subfloat[MNIST]{\includegraphics[width=0.45\textwidth]{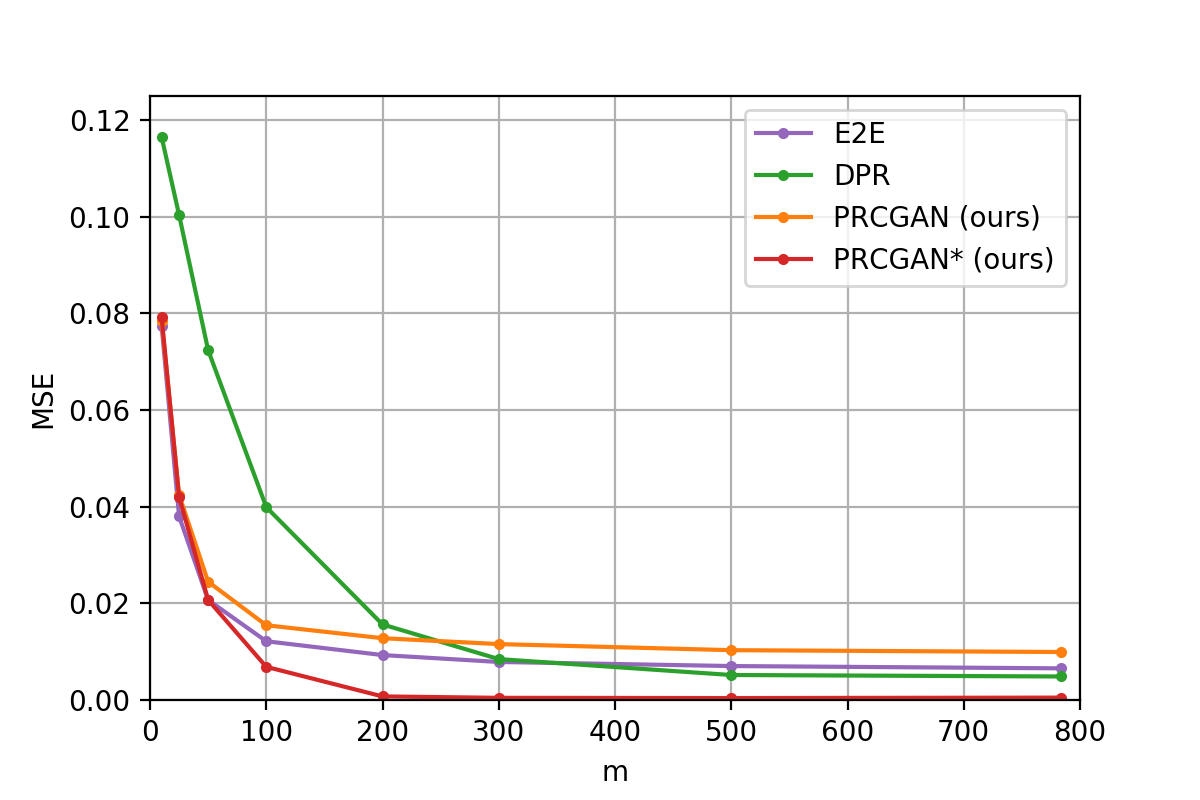}}\\
	\subfloat[Fashion-MNIST]{\includegraphics[width=0.45\textwidth]{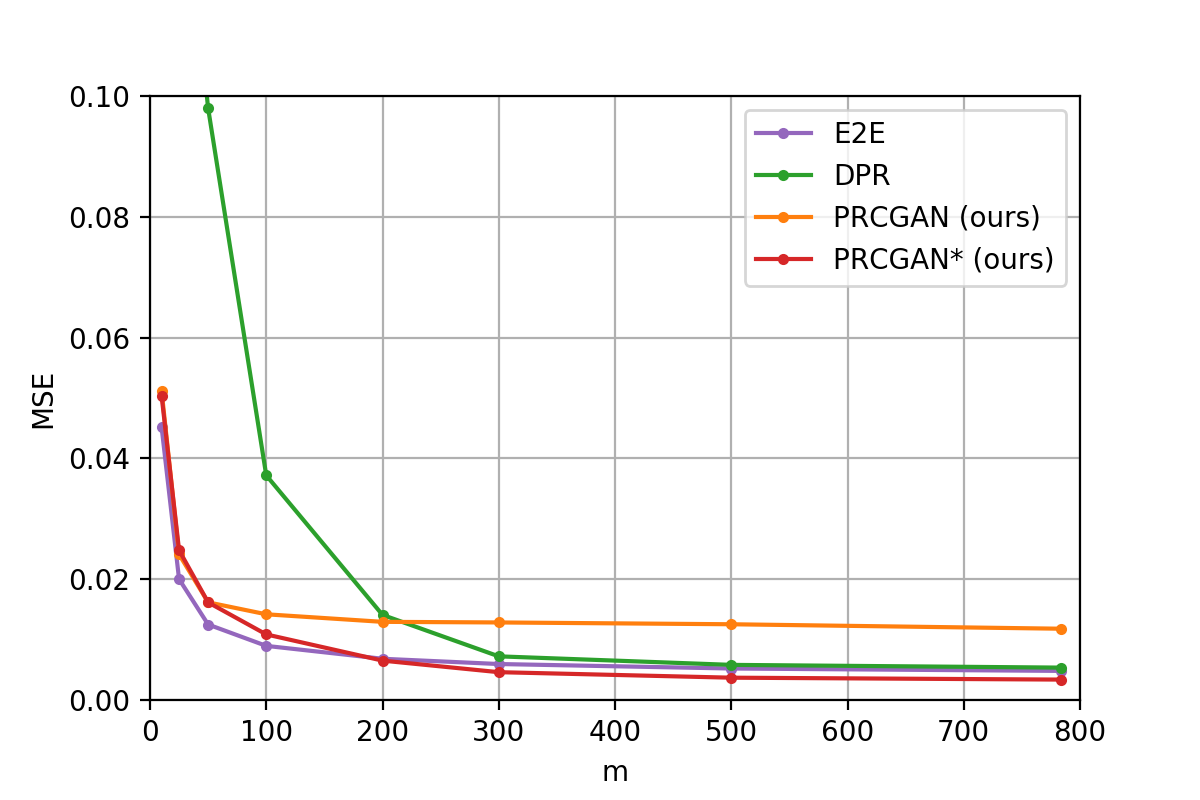}}
	\caption{Comparison of the PRCGAN, the E2E and generative prior approaches for different numbers of measurements $m$.}
	\label{fig:gaussianplots}
\end{figure}

For the next experiment we compared our PRCGAN approach to the DPR approach on the compressive phase retrieval problem with different amounts of information available.
For measurements we took a Gaussian measurement matrix $A \in \mathbb{R}^{m\times n}$, where $A$ has random entries sampled from $\mathcal{N}(0, 1/m)$.
While for Fourier phase retrieval with a fixed measurement matrix $\mathcal{F}$ our only choice to conduct a harder experiment was to use harder dataset conditions, we can now smoothly adjust the hardness of reconstruction by increasing the number of measurements $m$.
Knowledge of the measurement matrix $A$, that we kept fixed for each choice of $m$, allowed us to use the measurements $y$ also for the training of our PRCGAN model. 
For this experiment we took the same MNIST and Fashion-MNIST datasets and the same model structures for the PRCGAN and VAE as described in Section \ref{MNIST-Fashion}.
Since for MNIST a latent space of $20$ worked better for the VAE in the Gaussian measurement case, this was our only replacement to the model structure.
Figure \ref{fig:gaussianplots} shows the results for different values of $m$, where the maximum $m$ was chosen as $m = n$.
Higher values of $m$ are possible, but we do not expect better results, since Figure \ref{fig:gaussianplots} shows that there is barely a difference between the reconstruction error for $500$ and $784$ measurements.
As an evaluation metric we chose the MSE.
Since the Gaussian measurements are not invariant under translation or $180$ degree rotations as for the Fourier measurements, no image registration was needed for this experiment.
First, we report the results for the DPR approach, where we optimized the latent space of the VAE to minimize the measurement error.
Next, we evaluated the PRCGAN and E2E approach, where we trained a new model for each value of $m$.
For PRCGAN* we additionally optimized the latent variable input for the PRCGAN models.
Both plots show that lowering the measurement size increased the reconstruction error for all approaches.
Among both approaches with latent optimization our PRCGAN* outperforms the DPR approach for all sizes of measurements.

The reconstruction quality of DPR was strongly influenced by decreasing the number of measurements.
Figure \ref{fig:gaussian_measurements} shows the reconstructions of the different methods for a varying number of measurements $m$. 
Even for a quite small number of measurements, e.g., $m=200$, the PRCGAN* approach succeeded in reconstructing the lettering on the sweatshirt, while the other methods, in particular the E2E, fail to do so.

\section{Limitations}
\begin{figure}[H]
	\centering
	\includegraphics[width=0.4\textwidth]{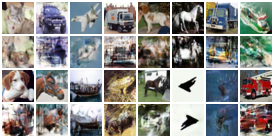}
	\caption{Original images of the CIFAR-10 dataset and unregistered reconstructions of our PRCGAN*. Odd rows contain original images, whereas even rows contain the reconstructions.}	
	\label{fig:limitations}
\end{figure}

We also evaluated our PRCGAN* on the CIFAR-10 dataset and observe that our approach fails on this dataset. This could be caused by the high variance of the images in the dataset. 
The results are shown in Figure \ref{fig:limitations}. 
We include this plot to briefly demonstrate the limitations of our method. 
The other methods did not perform better on this dataset.

\section{Conclusion}
In this paper, we show how conditional GANs can be employed to solve ill-posed phase retrieval problems. 
The proposed PRCGAN* combines the advantages of end-to-end learning and generative modeling and our experiments show that our method outperforms existing approaches in terms of quality.
Our conditional GAN approach yields smaller errors while still providing sharp images unlike the end-to-end approach.
Even when being trained on noise-free, synthetic measurements our model is still robust to noise.
The drawback of the PRCGAN* approach is that changing the measurement matrix requires retraining the model like it is the case for end-to-end learning.
However, at test time the computational cost of our model is the same as the cost of the DPR approach.





\bibliographystyle{IEEEtran}
\IEEEtriggeratref{20}
\bibliography{bibliography}
%

\end{document}